\documentclass[]{mn2e}

\usepackage{amssymb}
\usepackage{graphicx}
\usepackage{amsmath}
\usepackage{graphics}

\usepackage{color}

\title[IR plateau and continuum emission]{On the carbonaceous carriers of IR plateau and continuum emission}
\author[R. Papoular]{R. Papoular$^{1}$\thanks{E-mail:
papoular@wanadoo.fr}\\
$^{1}$Service d'Astrophysique and Service de Chimie Moleculaire,\\
CEA Saclay, 91191 Gif-s-Yvette, France}

\begin{document}

   \maketitle
\label{firstpage}

\begin{abstract}
This study explores the molecular origins of plateaus and continuum underlying IR and FIR  bands emitted by compact nebulae, especially proto-planetary nebulae. Computational organic chemistry codes are used to deliver the vibrational integrated band intensities of various large, typical carbonaceous structures. These spectra are composed of a rather continuous distribution of weak modes from which emerge the fingerprints. The 6 to 18-$\mu$m region is interspersed with a great many weak lines, to which the plateaus are assigned. Similarly, the far IR spectrum is ascribed to the phonon (skeletal) spectrum which is readily identified beyond 18 $\mu$m.

The absorptivities and absorption cross-sections per interstellar H atom deduced from these spectra are comparable with those of laboratory dust analogs and astronomical measurements, respectively. Moreover, the 5-35 $\mu$m spectra of two typical proto-planetary nebula were reasonably well simulated with combinations of molecules containing functional groups which carry the 21- and 30-$\mu$m bands, and molecules devoid of these but carrying strong phonon spectra.

These results may help understand the emergence of plateaus, the origin of continua underlying FIR bands, as well as the composition of circumstellar dust.

\end{abstract}




\section{Introduction}
In a recent publication, Kwok and Zhang \cite{kwo11} analyzed the IR spectra of 3 typical nebulae: a PPN (Proto-Planetary nebula), a PN (Planetary Nebula) and an HII region. They found that a general and meaningful decomposition of these spectra can be made into (atomic) lines, 
(molecular) bands, plateaus (underlying bands in spectral regions where their density is high) and one or more continua underlying all. Following Kahanp$\ddot{\rm a}\ddot{\rm a}$ et al. \cite{kah}, they suggested, on the basis of their analysis of spectral archives, that plateaus and continua might somehow be linked to bands through a physical relationship. They went on to surmise that amorphous organic solids with mixed aromatic-aliphatic structure are good candidate carriers.

Many UIB (Unidentified Infrared Band) sources are also known to emit an underlying continuum starting around 6 $\mu$m and rising beyond (see, for instance, Smith et al. 2007), which cannot be assigned to the central star nor to thermal silicate dust. 

\bf The purpose of the present work is to determine the nature of the relationship, if any, between bands, plateaus and continuum. More precisely we seek to determine if the underlying plateaus and continuum, or part of it, can be due to the same structure that carries the bands. 
\rm Kwok et al. \cite{kwo01} addressed the plateau problem long ago, and attributed this phenomenon to a complex collection of alkane and alkene side chains. It is argued below that aliphatic carriers are not the sole origin of plateaus, which also occur in more ``aromatic environments", like PNe and PDRs (see Kwok and Zhang 2011).

There is a natural tendency to associate preferentially lines and bands with atoms and molecules, on one side, and on the other, continuum with solid grains. But where is the limit between grains and large molecules ? 
\bf Li and Draine \cite{li12} aptly suggested that large molecules, ultra small grains and very small grains be considered as synonymous terms. They also set an upper size limit at 0.025 $\mu$m for ``very small grains" (including PAHs) for which they computed stochastic heating, larger sizes being the domain of ``grains" (including ``classical grains"). \rm
In space, however, dust grains evolve into atoms and molecules and vice versa, depending on the environment. Besides, the carriers of the UIBs are known to be relatively large structures essentially made of carbon and hydrogen, while the underlying continuum, too, is most often attributed to some form of hydrogenated carbon 
(see, for instance, Rouleau and Martin 1991). It is therefore likely that a more or less smooth transition occurs between these types of spectra, being driven by continuous grain size variation. 

The fingerprint spectral region of hydrocarbon molecules extends from 3 to about 13 $\mu$m. When O, N and S atoms are included in the structure as in kerogens and CHONS, other specific bands emerge below and above that range, especially strengthened, around 30 $\mu$m, by the presence of Oxygen (see Papoular 2000, Grishko and Duley 2002). On the other hand, the theory of solids teaches that their continuum is associated with \it phonons \rm\, (also called \it skeletal modes\rm\, in molecular parlance) which extend all over the structure, by contrast with the fingerprint features, whose vibrations are spatially limited to chemical functional groups. According to Debye (see Rosenberg 1988), the upper limit of the phonon frequency in crystals is
 \begin{equation}
\omega_{ph}=(6\pi^{2}N)^{1/3}v\,\,
\end{equation}
where $N$ is the atomic density of the solid, and $v$, the sound velocity. Typically, on Earth, $N=10^{24}$ at.cm$^{-3}$, and $v=2000$ m.s$^{-1}$, leading to a Debye wavelength of $\sim20\,\mu$m, roughly defining a lower limit to the phonon spectrum. Little or no phonon continuum is therefore to be expected below this limit. For a finite sample, the higher wavelength limit is set by the size and shape of the sample and increases with its size, reaching into the millimeter range.

Moreover, Debye's elementary theory also shows that the spectral density of phonons increases as the square of their frequency, i.e. maximum density occurs at Debye's wavelength. The phonon continuum should therefore overlap prominent PPN bands (21 and 30-35 $\mu$m) and perhaps even the redder UIBs. 

As shown in the present work, these general results are confirmed by the modeling of large and complex molecules and, so, evoke the possibility that the feature carriers themselves might also be responsible for at least part of the underlying continuum. This, in turn, prompts the more general question of the composition of the continuum carriers.

 In the case of oxygen-rich stellar envelopes, no attempt was ever made to distinguish the 10- and 18-$\mu$m peaks from their ``underlying continuum": they are all fitted together with measurements on model carriers.

In C-rich environments, on the other hand, it is usually agreed that the continuum is carried by some form of hydrogenated amorphous carbon (HAC), while the 11.3-$\mu$m feature, if any, is due to some form of silicon carbide, SiC, both in solid grain form. The question then arises whether the HAC assignment is not too restrictive, as it excludes the likely presence of heteroatoms. In the same vein, one may wonder if nano-sized structures may contribute to the continuum, as do solid (micro-sized) grains, and if so, what their temperature should be to fit observed spectra.

The problem of the plateaus is different, as these are localized in the finger print regions, and are therefore more likely to be linked to each fingerprint.

On top of these issues, one may question the rationale behind the modeling of the underlying ``continuum" with black bodies having a monotonous emissivity ($\lambda^{-n}$); indeed, even some HAC materials exhibit FIR (Far Infra Red) features on top of their extended continuum (Grishko et al. 2001).

These issues are best illustrated in compact, luminous nebulae. Planetary nebulae are particularly interesting in this respect, as several were measured up to very long wavelengths (see for instance Zhang and Kwok 1991), thus providing an opportunity to revisit models of the continuum $per se$ and far beyond the fingerprints.

To tackle these problems, we consider below several carbonaceous structures of different sizes and compositions which will be treated by means of standard molecular modeling algorithms to deliver absorption and emission spectra, as described in Sec. 2. \bf No attempt is made here to model specific astronomical features in detail. The aim is rather to try and illustrate as best we can the emergence of plateaus and continua which is shown below to be largely independent, qualitatively speaking, of particular structures. While those examples of structures that were chosen here are strongly inspired by the kerogen model of UIB carriers developed by Papoular \cite{pap01}, \cite{pap10} and \cite{pap12}, much more aromatic carriers, like those proposed by Li and Draine \cite{li12}, or mixtures of aliphatic and aromatic organic nanoparticles (MAONs proposed by Kwok and Zhang 2011) would fit as well.\rm

Obviously, electronic absorption is not taken into account in this study, which is only concerned with atomic and molecular vibrations.

The paper is organized as follows. Absorption spectra of simple and pure hydrocarbon chains shown in Sec. 3 illustrate the formation of a  phonon spectrum, clearly separated from the usual CH bands, as the size increases. An HAC model is built in Sec. 4 by interlinking branched (ramified) chains, increasing the number of sp$^{3}$ bonds, and adding a small amount of oxygen; this blurs the boundary between fingerprints and phonons, and gives rise to features within the phonon range. Finally, Sec. 5 discusses structures that were invoked to model UIBs and PPN features by analogy with the constitutive components of kerogen, as suggested by Papoular \cite{pap01a}, \cite{pap11} and Kwok and Zhang \cite{kwo11}: short, branched or oxygen-bridged alkane chains (-CH$_{2}$-), naphtenic chains (linear chains of benzenic rings) and small, compact, benzenic ring clusters interconnected with the former. The phonon spectrum is evidenced in all cases and may extend to several thousands of micrometers in wavelength. In all cases, bands are found to coexist with plateaus or continuum.

In order to ascertain the relevance of these model outcomes to the present astronomical issues, in Sec. 6, the computed spectral intensities are quantitatively compared, on the one hand with the absorptivities measured on various laboratory model candidates, and on the other hand, with the astronomical absorption cross-sections per H atom measured towards various solar environments.

Finally, in Sec. 7, the emission spectra of our model carriers are computed for the case of  radiative equilibrium in a strong radiation field, as in compact nebulae. Fits to a couple of astronomical spectra are displayed and the best-fit carrier number densities and temperatures are discussed.

Consequences of these findings are discussed in the Conclusion.

\section{Computational methods}

When studying emission from dust (or big molecules), at least two opposite generic circumstances must be considered, namely strong and weak excitation by ambient radiation (or any other process). In the latter case, which applies mainly to the ISM (InterStellar Medium), the interval between two successive excitation events is so long that the dust particle returns to its ground state in between; this case is outside the scope of the present study, as we are interested here in compact nebulae. On the other hand, in the first case, energy accumulates in the particle until 
it reaches radiative equilibrium, say at temperature $T_{d}$; then, the spectral radiance at wavelength $\lambda$ is that of a gray body,
 
\begin{equation}
I_{\lambda}=\alpha(\lambda).d.BB(T_{d},\lambda),
\end{equation}

where $\alpha$ is the absorptivity of the material, in cm$^{-1}$, $d$ its depth in cm along the line of sight, and $BB$ the black body law; the optical thickness is assumed much smaller than 1. For a solid, $\alpha$ is a continuous function of the solid's dielectric functions or extinction coefficient, which are also continuous. Molecular spectra, on the other hand, are discrete line spectra, and $\alpha$ \it for a single line \rm is deduced from the corresponding integrated band absorption, $A$, by

\begin{equation}
\alpha(\lambda)=10^{2}A(\lambda)\frac{C}{\Delta\nu}\,\,,
\end{equation}

where $\alpha(\lambda)$ is in cm$^{-1}$, $\Delta\nu$ is the band width (cm$^{-1}$) and $C$ is the molecular density (mol.l$^{-1}$) along the line of sight. For one type of structure, Equation 2 then translates into 

\begin{equation}
I_{\lambda}=10^{2}A(\lambda)\frac{LC}{\Delta\nu}\,\,BB(T_{d},\lambda),
\end{equation}
 
where $L$ is the thickness of the emitting cloud.

\it To approach a continuous spectrum, it is necessary to consider either very large structures, or a large number of smaller, non-identical structures. The latter is computationally more tractable and will be adopted here. \rm

 The implementation of the above considerations is based on the use of various algorithms of computational organic chemistry, as embodied in the Hyperchem software provided
by Hypercube, Inc., and described in detail in their publication HC50-00-03-00, and cursorily, for astrophysical purposes, in Papoular \cite{pap01b}. Here I use the
 improved version Hyper 7.5. While almost all the computations used standard algorithms, some required writing special purpose subroutines, a 
capability also provided by the software. Semi-empirical methods provided by the software (PM3 and AM1, acronyms for Prametrized Model No 3 and Austin Model No 1) were used for molecules up to about 500 atoms in size; their accuracy for spectral frequencies is a few percent. Anharmonicity is built in the algorithms, in terms of parameters which are tailored by comparison with laboratory experiments on hydrocarbon molecules. 

Details of the computation procedure can be found in Papoular \cite{pap01b}. Briefly, a structure is first sketched on the computer screen. Its geometry is then automatically optimized by minimizing the total potential energy. With semi-empirical simulation methods, it is then possible to launch a Normal Mode Analysis which delivers the spectrum of integrated absorption bands, $A(\nu)$, at frequency $\nu$. A very useful subroutine allows one to study the characters of each mode individually by displaying on the screen the velocity vector of each atom of the molecule. This velocity field is a powerful tool for distinguishing fingerprints from plateau or continuum.

\section{Pure hydrocarbon chains}

\begin{figure}
\resizebox{\hsize}{!}{\includegraphics{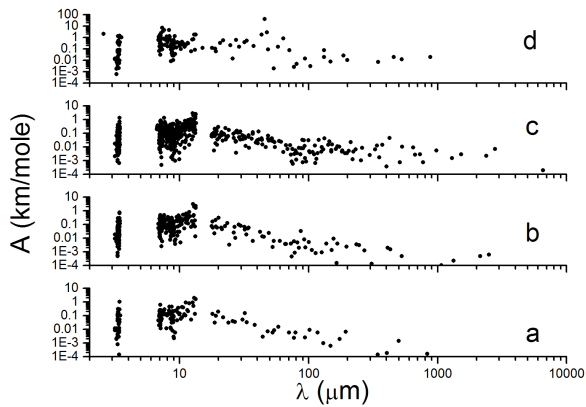}}
\caption[]{The integrated absorption band intensity spectra of a few representative aliphatic chain structures. From bottom to top: a) Chain 1, with 53 atoms (17 C, 36 H); b) Chain 2, with 106 atoms (34 C, 72 H); c) Chain 5, with 263 atoms (87 C, 176 H); d)Chain 1b, with 54 atoms (17 C, 36 H, 1 O). Each point represents a vibration of the system. The scatter presentation was preferred to the usual column graphs because it highlights the vertical dispersion, and the decreasing density of bands with increasing wavelength. In (a),(b) and (c), there is a clear separation between the CH stretching and the other fingerprint region (6-13 $\mu$m), and between the latter and the phonon continuum. The insertion of only one O atom considerably increases the intensities, blurs the lower edge of the phonon region, and gives rise to a new band around 40 $\mu$m. The vertical dispersion is much larger in the fingerprint region than in the phonon region.} 
\label{Fig:specchalks}
\end{figure}

A very large number of typical structures were previously analyzed (Papoular 2011), from which some can be selected for the present purposes. If the continuum properties are to be properly highlighted, it is better to start with the simplest possible structure so as to avoid unwanted finger-prints. An example of this is the  alkane chain (-CH$_{2}$-), a pure hydrocarbon, which is known to be a major constituent of kerogens (see Behar and Vandenbroucke 1986). The only fingerprints are the CH stretching ($\sim3.4\,\mu$m) and CH in plane and out-of-plane  (7 to 13 $\mu$m) vibrations, distinctly separate from the phonon modes. This is illustrated in Fig. \ref{Fig:specchalks} which plots the normal mode spectra of Chain 1,  with 53 atoms (17 C, 36 H), Chain 2, obtained by concatenating 2 identical copies of Chain 1, and Chain 5, obtained by concatenating 2 identical copies of Chain 2, with 263 atoms (87 C, 176 H). Two other structures were also analyzed (but are not represented in Fig. \ref{Fig:specchalks}): Chain 3,  with 206 atoms (68 C, 138 H) and Chain 4, with 257 atoms (85 C, 172 H).

 As predicted, the phonon spectrum begins near 18 $\mu$m;  it is clearly separated from the finger prints below; its upper limit increases almost linearly with the chain length. Also, the spectral density of phonons decreases steeply as $\lambda$ increases. For all 3 structures, the relative number of phonon vibrations is about 21 $\%$. In general, \it these differ from fingerprint modes in that their intensity is weaker, and that each mode extends over the whole structure. \rm
 \begin{figure}
\resizebox{\hsize}{!}{\includegraphics{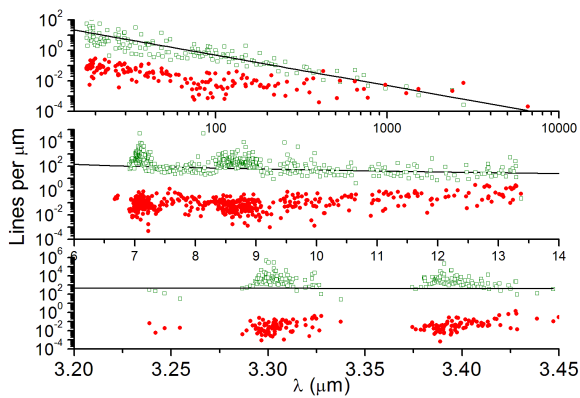}}
\caption[]{The particular case of Chain 5 of Fig. \ref{Fig:specchalks}. $red \,dots$: integrated absorption band intensities; $green \,open\, squares$: spectral density of vibrational modes; $black \,line$: Debye's power law in $\lambda^{-2}$. From bottom to top, the near-, mid- and far infrared domains. The larger dispersions of intensities and densities in the UIB windows give rise to plateaus. Note the extension of the FIR continuum up to 7000 $\mu$m. Color on line}
\label{Fig:specchalk6}
\end{figure}

 Figure \ref{Fig:specchalk6} elaborates on these spectral properties in the typical case of Chain 5. Both the density and the intensity of the modes are represented in 3 generic spectral windows. The straight line encompassing all 3 windows is a power law in $\lambda^{-2}$ which  illustrates Debye's law. Remarkably, within the near- and mid-IR bands, the mode densities are well above this line. The shape of a particular band (for instance, the 3.3-$\mu$m band) is determined both by the mode densities, which peaks roughly in the middle, and the mode intensities, which are more dispersed near the middle of the band.

Note the strong dispersion of band intensities, even within a small range of wavelengths.  Both within the CH stretchings and the MIR (Mid Infra Red)) fingerprint regions, very weak and very strong bands coexist. But none is observed between 3.6 and 6 $\mu$m, and 13.5 and 17 $\mu$m, where there are no phonons. What happens is made clear by visualizing the atomic velocity vectors on the screen: two very close modes of vibrations, both involving the same type of functional group, can have so different topologies that, in one, the vibrational dipole moments add up constructively as they do for a single well defined chemical group, while, in the other, they compensate each other more or less. The observed closeness of very strong and very weak lines is characteristic of finite and disorder systems, (but forbidden by symmetry in atomic crystals).

\begin{figure}
\resizebox{\hsize}{!}{\includegraphics{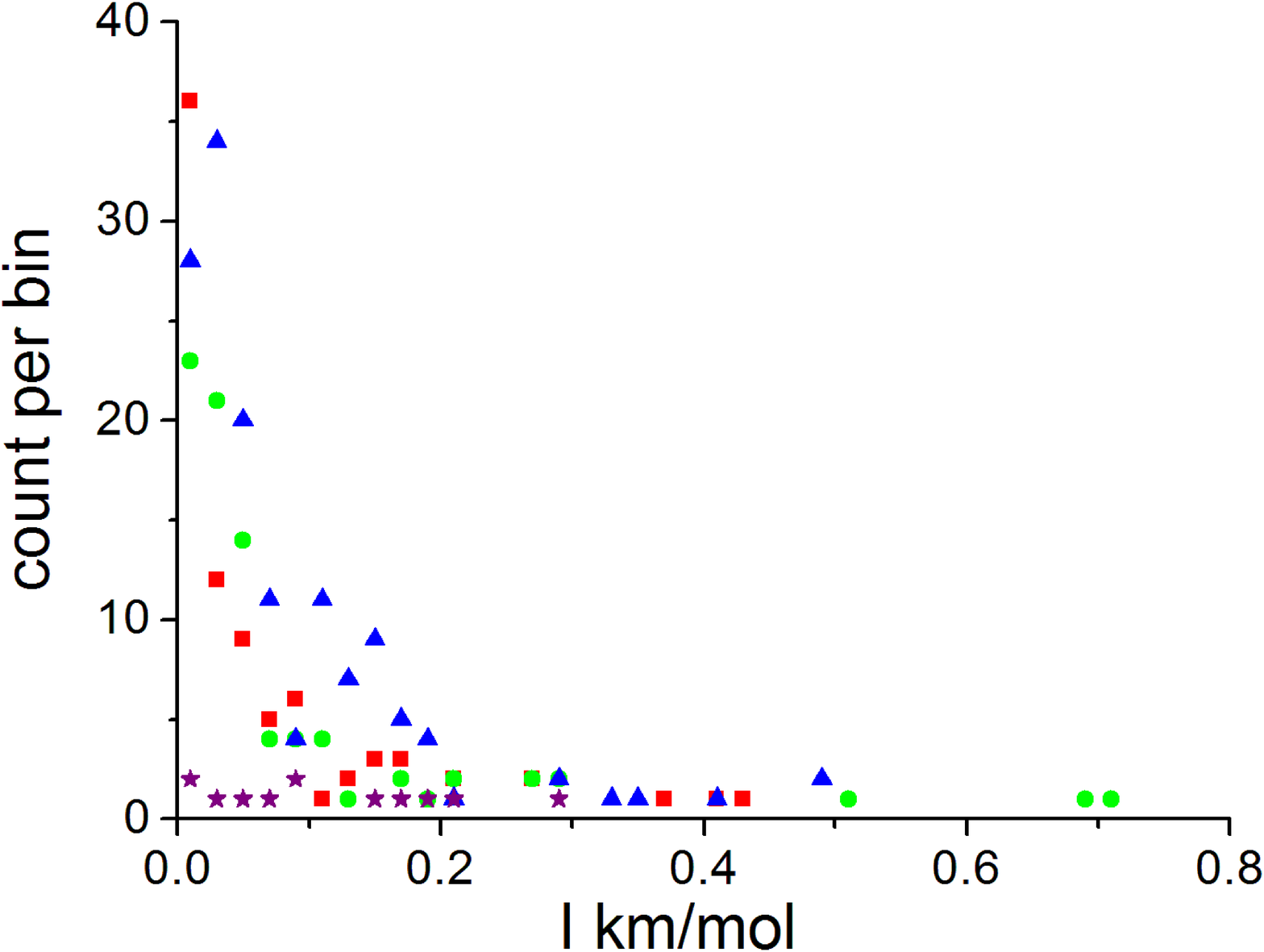}}
\caption[]{Histograms of the integrated intensities within each of 4 characteristic spectral windows around 3.3 (red squares), 3.4 (green dots), 8.6 (blue triangles) and 20 $\mu$m (purple stars), in the spectrum of Fig. \ref{Fig:specchalk6}. The abscissa represents successive 0.02 km/mol bins of mode intensity, while the ordinate plots the number of mode counts in each bin. 
 In the 3  fingerprint windows, the number of counts decreases steeply with increasing intensity; most of the counts correspond to weak intensities. In the phonon window, the histogram is almost flat and limited to weak intensities. Color on line.}
\label{Fig:counts}
\end{figure}

This phenomenon holds a clue to the formation of spectral plateaus. Figure \ref{Fig:counts} displays histograms of the integrated intensities within each of 4 characteristic spectral windows around 3.3, 3.4, 8.6 and 20 $\mu$m. The abscissa represents successive 0.02 km/mol bins of mode intensity, while the ordinate plots the number of mode counts in each bin. \it 
 In the 3  fingerprint windows, the number of counts decreases steeply with increasing intensity; most of the counts correspond to weak intensities. The latter form an almost continuous pedestal from which only a few distinct features protrude, giving the impression of fingerprints sitting upon a ``plateau".\rm \,This phenomenon is easily modeled mathematically by adding up a few strong, and many weak Lorentzian profiles, all with a finite width, within a finite interval of frequencies. The inverse process, disentangling the plateau from the protruding fingerprints, is much more difficult to carry out without preconceptions. A mathematical construction of a spectral plateau is only an approximate representation of the distribution of weak vibrational modes.

 \it In the phonon window, the histogram is almost flat and limited to weak intensities. These are responsible for the FIR continuum, as well as for the plateaus. Here, however, fingerprints are much less frequent.\rm To illustrate the emergence of FIR fingerprints, a Chain 1b was built from a Chain 1 by replacing one of its CH$_{2}$ groups, in the middle, by an O atom (O bridge), a substitution that is likely to occur in space in view of the abundance of the species involved. It is also known to occur  frequently in kerogens, and to enhance the intensity of many other vibrations of the same structure. Its spectrum is shown in Fig. \ref{Fig:specchalks}d. The finger prints now extend below 3 $\mu$m and above 13 $\mu$m, and far above the phonon edge, especially around 40 $\mu$m. Also, note the band intensities, higher by an order of magnitude, than those of the previous structures.

\it The comments made above on the coexistence and meaning of strong and much weaker vibrations within a small frequency range also apply to this structure, around 40 $\mu$m and, by extension, to the PPNe features. \rm

Even a minute alteration of a given structure (such as subtraction or addition of a -CH$_{2}$- group, or a slight bending of the chain) suffices to change all mode frequencies, albeit to a very small extent. None of the computed mode frequencies coincides with any other. A myriad of such structures along a line of sight may indeed mimic a continuum of frequencies, be it a band, a plateau or a phonon continuum. In real particles, this transition to a continuous spectrum is expedited by the anharmonicity of bonds, which couples neighboring modes and imparts a finite width to each. In the frame of the present treatment, approaching continuity will only become possible after enough similar, but not identical, spectra have been obtained.

 \it In any event, the emergence of embryonic ``plateaus" and continua in such simple structures makes one wonder if (all) the continuum underlying a band should be assigned to an altogether different material than the carrier of the band itself, and give it a temperature of its own. The two may be, at least partially, inseparable.\rm

\section{HAC-like structures}

\begin{figure}
\resizebox{\hsize}{!}{\includegraphics{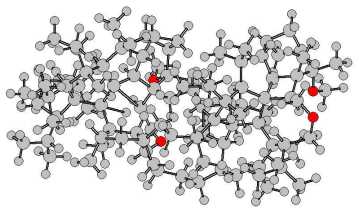}}
\caption[]{hac: a ramified, HAC-like, structure; 284 atoms; large gray balls: carbon (106), small gray: hydrogen (174), black: oxygen (4). Color on line.} 
\label{Fig:molhac5}
\end{figure}
\begin{figure}
\resizebox{\hsize}{!}{\includegraphics{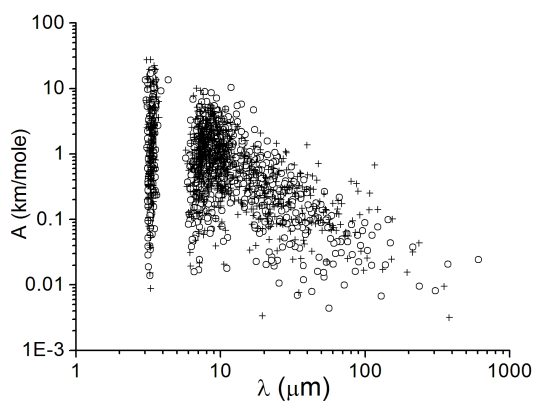}}
\caption[]{Integrated absorption intensity spectrum of two HAC-like structures. $open\, circles$: oxygen-free; $pluses$: 4 O-bridges included, as shown in Fig. \ref{Fig:molhac5}. Color on line.}
\label{Fig:spechacs}  
\end{figure}

HAC is a prime carrier candidate for the continuum. Several models of HAC have been proposed (see, for instance, Dischler at al. 1983; Robertson 1986; Walters and Newport 1995; Duley 1995). By contrast with the previous structures, these are ramified (cross-linked). Most of these models are designed for material science and do not include oxygen. In space, however, the inclusion of oxygen atoms in such structures is highly probable. One of such object studied in this work, including 4 oxygen bridges, was built along these lines; it is represented in Fig. \ref{Fig:molhac5}. Another HAC-like structure was derived from the previous one by replacing all 4 O atoms with C atoms. \bf These structures are not intended as new materials but, rather, as yet other models of laboratory HAC, used here on the same footing as other types of models to understand the origin of plateaus and continua. \rm

The spectra of both structures are shown in Fig. \ref{Fig:spechacs}. The gap observed between 14 and 17 $\mu$m in the previous spectra is now partially filled. This is due to the higher connectivities of C atoms in this structure, which couples mid-IR and far-IR modes and gives rise to mixed modes in the gap. Most importantly, the intensities are increased, on average, by a factor about 10. As was the case for Chain 1b (Fig. \ref{Fig:specchalks}), the insertion of oxygen gives rise to several additional high intensity modes from 30 to 100 $\mu$m.

 \section{CHONs or kerogen-like structures}
 
 \begin{figure}
\resizebox{\hsize}{!}{\includegraphics{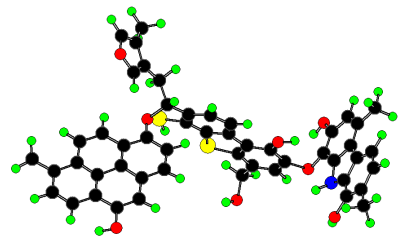}}
\caption[]{ker1: a component of a model kerogen structure; large black ball: carbon (51), small green: hydrogen (39) medium red: oxygen (8); blue: nitrogen (1), yellow: sulfur (2); total: 101 atoms. Note the presence of small aromatics and other carbon rings. Its vibrational spectrum is given in Fig. \ref{Fig:specSPs}a. Color on line.} 
\label{Fig:molSP138}
\end{figure}

 \begin{figure}
\resizebox{\hsize}{!}{\includegraphics{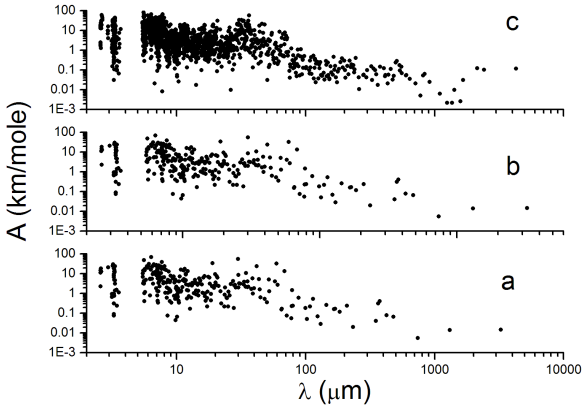}}
\caption[]{The integrated absorption band intensity spectra of kerogen-like structures: a)  ker1: the structure illustrated in Fig. \ref{Fig:molSP138} (101 atoms, 51 C, 39 H, 8 O, 1 N, 2 S); b) ker2: identical to (a), except for the substitution of an H atom for an SH group; c) ker3: a larger kerogen-like structure: 5 interlinked structures of the type illustrated in Fig. \ref{Fig:molSP138} (493 atoms, 254 C,125 H, 40 O, 5 N, 9 S).} 
\label{Fig:specSPs}
\end{figure}

More complicated and richer structures, containing some of the most abundant elements in the ISM (H, C, O, N and S), and called CHONS for this reason, have often been proposed as IS dust.
On Earth, kerogens are also known to carry these elements, arranged in aromatic structures as well as pentagons, as illustrated in Fig. \ref{Fig:molSP138}, inspired by the work of Speight \cite{spe}. The integrated intensity spectrum of this structure is shown in Fig. \ref{Fig:specSPs}a (ker1). Another, very similar structure was obtained by replacing just one SH group by an H atom (ker2). Its spectrum is plotted in Fig. \ref{Fig:specSPs}b.

A larger kerogen-like structure, ker3 (493 atoms), was built by interlinking 5 identical structures of the type drawn in Fig. \ref{Fig:molSP138} (101 atoms). This is the largest structure that could be analyzed with semi-empirical methods to give its normal mode vibrations (1 month computation time). Its spectrum is shown in Fig. \ref{Fig:specSPs}c.

Note that these spectra carry the strongest integrated intensities of all the structures studied here. This is an effect of the inclusion of aromatic rings and OH functional groups.
The near and mid-IR fingerprints are prominent in these spectra (see Fig. \ref{Fig:alpha} and Fig. \ref{Fig:sigma}).

\section{Comparison with laboratory and astronomical measurements}

For purposes of comparison with laboratory measurements and astronomical observations, the absorption intensity spectra have to be averaged, as they are with any spectrometer having a finite spectral resolution. This is a delicate operation because of their large dispersion mentioned above. It was effected here by first concatenating as many spectra of the same category as are available. Each concatenated spectrum was then broken into 3 intervals: 3-4, 5-17 and beyond 17 $\mu$m, and each was smoothed by adjacent averaging over N points. In each interval, N was chosen so as to smooth the spectrum without erasing the features. As noted earlier, smoothing is the lesser necessary, the larger the structure and the larger the number of concatenated spectra of similar structures.

 These averages were then converted into absorptivity curves using Eq. 3, with a uniform C-atom density of $8.3\,10^{22}$ cm$^{-3}$, typical of a solid.  As the radiative lifetime of IR vibrations is very long, the band width is more likely determined by the non-radiative lifetime, which, for present purposes, is the relaxation time constant for the vibrational dynamic equilibrium to settle after perturbation of the molecule, and is of the order of 1 ns (Papoular 2006), corresponding to $\Delta\nu=0.0333$\,cm$^{-1}$. The resulting conversion formula is

$\alpha=4.2\,10^{5}\frac{A}{N_{C}}$,

where $N_{C}$ is the number of carbon atoms in the selected structure.

\begin{figure}
\resizebox{\hsize}{!}{\includegraphics{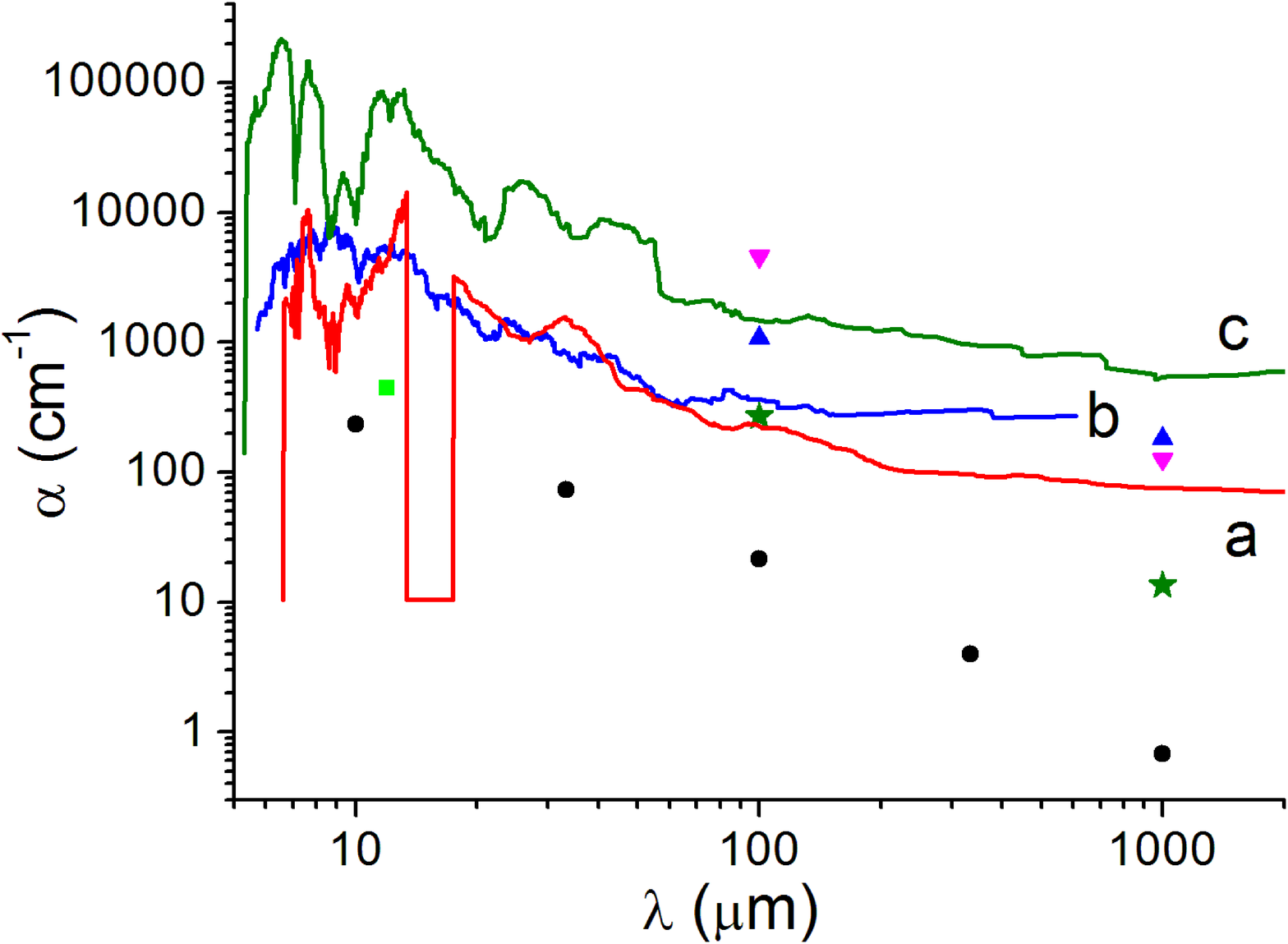}}
\caption[]{ The smoothed absorptivities of a)the 3 alkane chains of Fig. 1a,b,c (concatenated spectra), b)of the HAC-like  structures of Fig. \ref{Fig:spechacs} (concatenated), and c)of the kerogen-like structures of Fig. \ref{Fig:specSPs} (concatenated). Also plotted are laboratory measurements on black polyethylene by Blea et al. \cite{ble}, containing only 3.3 \% by weight of carbon-black microparticles (black dots); on a-C:H by Dischler et al. \cite{disB} (green square); on graphite (purple nablas), amorphous carbon, BE (blue triangles) and coal (olive stars), as reported by Mennella et al. 1995. Color on line.} 
\label{Fig:alpha}
\end{figure}

The results are shown in Fig. \ref{Fig:alpha} for the 3 alkane chains of Fig. 1a,b,c (concatenated spectra), for  the HAC-like  structures of Fig. \ref{Fig:spechacs} (concatenated), and for the kerogen-like structures of Fig. \ref{Fig:specSPs} (concatenated).

 Also plotted in Fig. \ref{Fig:alpha} are representative results of measurements on black polyethylene (charged with carbon-black pigments at 3 \% by weight) by Blea et al. \cite{ble}, on a-C:H by Dischler et al. \cite{disB}, and on various carbonaceous grains by Mennella et al. \cite{men}. The values obtained by the latter authors are given in cm$^{2}$g$^{-1}$, so they had to be converted into cm$^{-1}$ by multiplying them with the corresponding densities in cm$^{-3}$: 1.8 for benzene burning HAC BE, 1.5 for Mericourt coal and 2.1 for graphite.
 
 Observe again that the presence of benzenic rings and OH groups in the model structures of Fig. \ref{Fig:specSPs} enhances absorbance by a factor about 10 over that of the pure hydrocarbons of Fig. \ref{Fig:spechacs}.

\begin{figure}
\resizebox{\hsize}{!}{\includegraphics{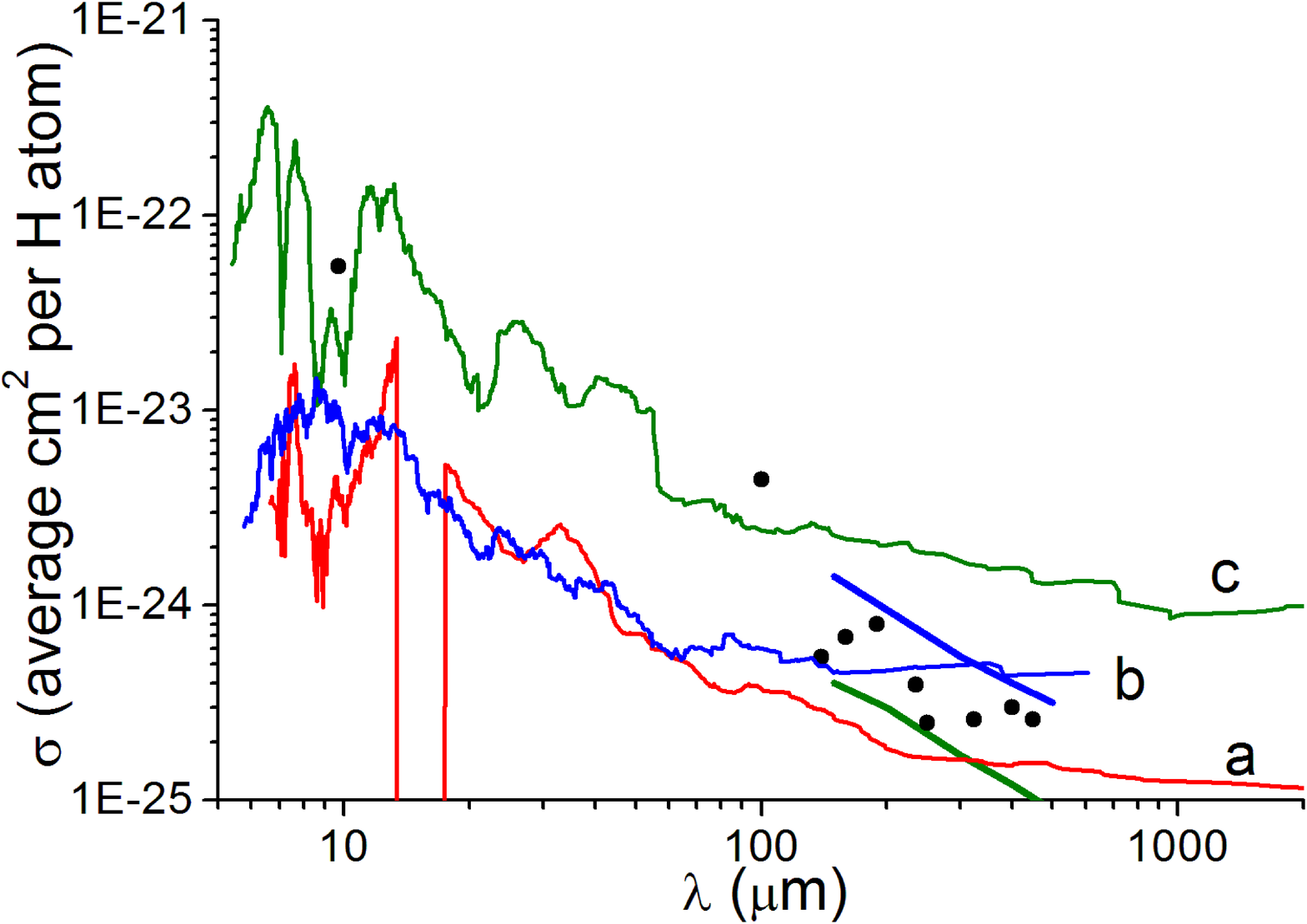}}
\caption[]{Cross-sections per H atom of IS space for a)the 3 alkane chains of Fig. 1a,b,c (concatenated spectra), b)of the HAC-like  structures of Fig. \ref{Fig:spechacs} (concatenated), and c)of the kerogen-like structures of Fig. \ref{Fig:specSPs} (concatenated). 
 For comparison, c-s's measured by Mezger et al. \cite{mez} towards compact and extended HII regions (blue and wine thick line segments). Also plotted as black dots, are c-s's deduced by various authors from absorption measurements towards solar neighborhoods, as compiled by Mezger et al. in their Table A1. Color on line.}
\label{Fig:sigma}
\end{figure}

We seek now to compare our simulation results with the dust cross-sections per H atom deduced from observations by Mezger et al. \cite{mez}. For this purpose, we note that the absorption c-s (cross-section) of a molecule is linked to its integrated band absorption by 

\begin{equation}
\sigma(cm^{2})=\frac{\alpha}{n}=\frac{A(km/mole)}{6\,10^{18}\Delta\nu(cm^{-1})}
\end{equation}

where $n$ is the density of molecules in cm$^{-3}$. Assume also that the mass ratio of dust to hydrogen gas is 1/300, half of which is in carbonaceous dust, the latter entirely consisting of alkane chains. The c-s per H atom is then given by

\begin{equation}
\sigma(cm^{2}/H at)=\frac{A(km/mole)}{N_{C}\cdot6\,10^{18}\cdot300\cdot2\cdot12\cdot0.0333}=\frac{A(km/mole)}{1.44\,10^{21}\,N_{C}},
\end{equation}
 This is plotted in Fig.\ref{Fig:sigma} for the 3 alkane chains of Fig. 1a,b,c (concatenated spectra), for  the HAC-like  structures of Fig. \ref{Fig:spechacs} (concatenated), and for the kerogen-like structures of Fig. \ref{Fig:specSPs} (concatenated). For comparison, c-s's measured by Mezger et al. \cite{mez} towards compact and extended HII regions are plotted on the same graph. Also included, are c-s's deduced by various authors from absorption measurements towards solar neighborhoods, as compiled by Mezger et al. in their Table A1. These results are confirmed, beyond 100 $\mu$m, by data from the Planck satellite (see, e.g. Abergel et al. \cite{abe11}.  

It appears from Fig. \ref{Fig:alpha} and \ref{Fig:sigma} that the absorptivities of kerogen-like model materials, used elsewhere to mimic the UIBs as well as the PPNe far-infrared features, are strong enough that they may also carry a non-negligible fraction of the observed astronomical FIR continuum, too. \bf Recall that $\sigma$ in eq. 6 scales like the mass ratio of dust to hydrogen gas. According to recent estimates, the value assumed here for this ratio, 1/300, is only a lower bound. The resulting values of $\sigma$ are, therefore, conservative estimates.  \rm

In both figures, the shape of the curves between 10 and 100 $\mu$m is roughly a power law with exponent about 1.5. The leveling off beyond 100 $\mu$m is disputable as the density of modes is too weak in that range. Computations of the emission of these structures by monitoring the variations of their dipole moments under perturbation (as elaborated by Papoular 2012) point, rather, to a nearly constant slope down to 1000 $\mu$m and beyond.  

Note that, while graphite is a very efficient continuum carrier, it can hardly be invoked in the present instance, for it has never been identified, nor is it likely to form, in circumstellar environments. 

\section{Emission spectra}

\begin{figure}
\resizebox{\hsize}{!}{\includegraphics{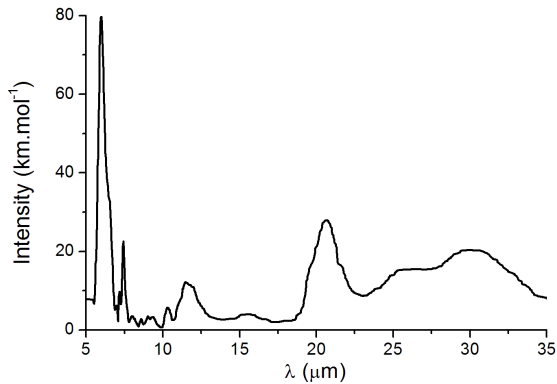}}
\caption[]{The smoothed integrated intensity spectrum resulting from the concatenation of the spectral lines of thiourea and thiourea derivatives with compact aromatics, thiourea derivatives with linear aromatics and aliphatic chains with attached OH chemical groups. The 21-$\mu$m feature peaks at 20.7$\mu$m; FWHM=2.3 $\mu$m. Note the shoulder at 26 $\mu$, reminiscent of the 26-$\mu$m spectral component conjectured by Kwok and collaborators. Adapted from Fig. 11 in Papoular \cite{pap11}.}
\label{Fig:thiourea}
\end{figure}

We have still to confront the model emission spectra with astronomical ones. For this purpose, we select two spectra of PPN (IRAS 223304+6147 and 22574+6609), exhibiting prominent 21 and 30 $\mu$m features, which were obtained by Zhang et al. \cite{zha10} with the IR spectrometer embarked on the \emph{Spitzer} satellite. A tentative fit to the conspicuous 21-$\mu$m bands  in these spectra was previously obtained by Papoular \cite{pap11} with a carrier model featuring the thiourea functional group (SC(NH$_{2})_{2}$) associated with various carbonaceous structures (mainly compact and linear aromatic clusters). In the same study it was also found that the equally conspicuous 30-$\mu$m band could be modeled by a combination of aliphatic chains made of CH$_{2}$ groups, oxygen bridges and OH functional groups. The concatenation of the spectra of all these structures gave the band spectrum reproduced in Fig. \ref{Fig:thiourea}, from Fig. 11 of the cited paper.

In order to fit the spectrum of a PPN, it was necessary to associate the band emissions with the emissions of two gray bodies having different temperatures, and absorptivities scaling like $\lambda^{-2}$. The only difference between the procedure followed in the paper just cited and the one used here is that, here, the gray bodies are replaced by tailored combinations of the 3 groups of structures described in Sec. 3, 4 and 5 above, and designated hereafter by ``dust" or ``continuum model carriers": a) the 3 alkane chains of Fig. 1a,b,c (concatenated spectra), b)  the HAC-like  structures of Fig. \ref{Fig:spechacs} (concatenated), and c) the kerogen-like structures of Fig. \ref{Fig:specSPs} (concatenated). In fact, it appeared, ultimately, that the alkane chains did not help because of a strong peak near 13 $\mu$m. Neither did the kerogens, because of strong spectral massifs from 25 to 50 $\mu$m; these are due to out-of-plane vibrations of the clusters of benzenic rings. Only the HAC-like structures were therefore retained. It was also found that at least 2 or 3 different dust temperatures were necessary: warm, cool and cold. The model carriers of the 21 and 30 $\mu$m bands were taken from Papoular\cite{pap11} and their concatenated spectrum is that of Fig. \ref{Fig:thiourea}. For the two selected PPNe, Table 1 details the temperatures and relative number densities of band and continuum model carriers. The relative compositions of the latter are the same for all 2 (or 3) temperatures. The standard deviations between spectrum and fit are, respectively, 0.18 and 0.1 (in the same arbitrary emission units as the ordinates) i.e. less than 10 \% of the peak emission.

\begin{table*}
\caption[]{Temperatures and relative compositions of\\
band and continuum model carriers for 1) IRAS 223304+6147\\
and 2) IRAS 22574+6609. Temperatures in Kelvin.}
\begin{flushleft}
\begin{tabular}{lll}
\hline
Name & 1 & 2\\
\hline
T(band)& 170 & 160\\
\hline
T(warm) & 170 & 160\\ 
\hline
T(cool) & 80 & 110\\
\hline
T(cold) & - & 60\\
\hline
Band & 0.018 & 0.003\\
\hline
Warm dust & 0.008 & 0.002\\ 
\hline
Cool dust & 0.974 & 0.019\\
\hline
Cold dust & - & 0.976\\
\hline
\end{tabular}
\end{flushleft}
\end{table*}

The fact that up to 3 different dust temperatures had to be invoked while only one temperature was used for the band carriers clearly shows that the dust carriers extend farther from the central star than the band carriers, and that their temperature is therefore distributed continuously over the radial distance.

Obviously, the fits can be improved by tailoring separately the components of dust and band carriers. In particular, the model carriers of the 21-$\mu$m and 30-$\mu$m bands are quite distinct (Papoular 2011); other astronomical spectra may therefore require different relative contributions from these components.

\begin{figure}
\resizebox{\hsize}{!}{\includegraphics{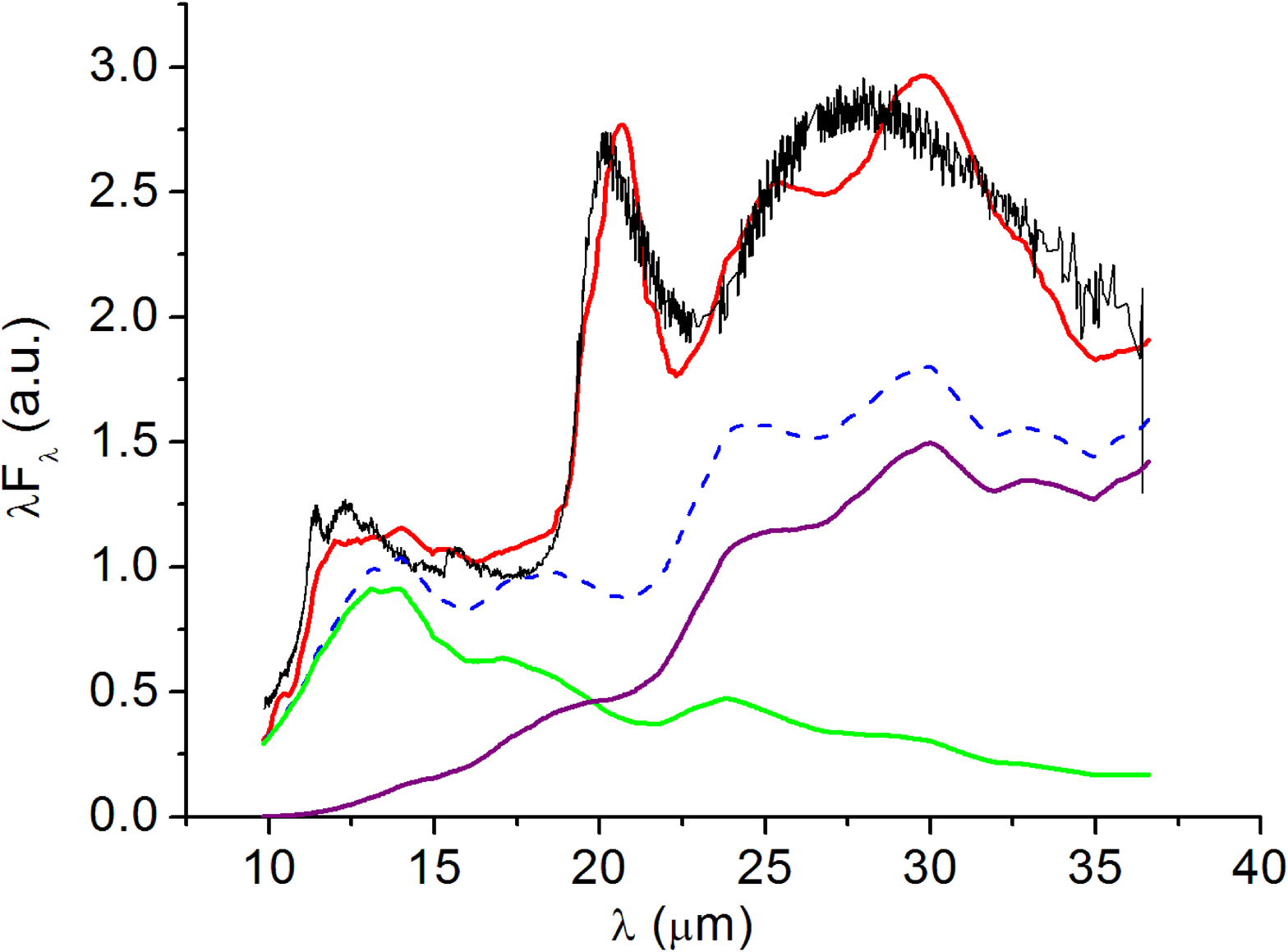}}
\caption[]{$Noisy \,curve$: recorded spectrum of IRAS 23304+6147 (Zhang et al. 2010); $continuous \,line$: tentative fit with a synthetic spectrum (see text); $dashed \,line$: contribution of ``continuum" carriers, as opposed to ``band" carriers i.e. those carrying specifically the 21 and 30 $\mu$m bands (see Table 1). The spectra of the warm (170 K) and cool (80 K) components of dust are drawn in green and purple lines, respectively. The standard deviation between spectrum and fit is 0.18 (in the same arbitrary emission units as the ordinates) i.e. less than 10 \% of the peak emission. Color on line.}
\label{Fig:iras1} 
\end{figure}

\begin{figure}
\resizebox{\hsize}{!}{\includegraphics{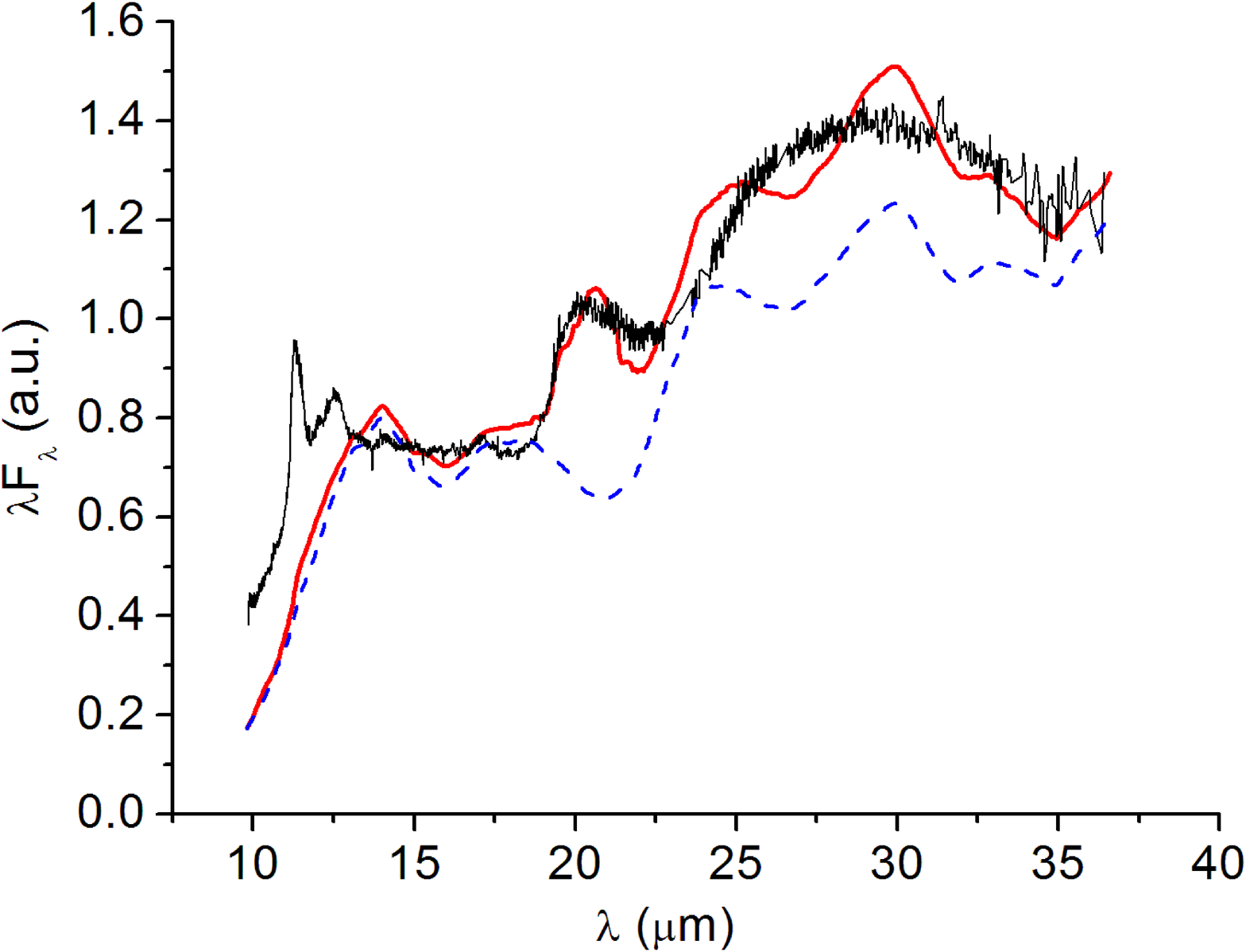}}
\caption[]{$Noisy\, curve$: recorded spectrum of IRAS 22574+6609 (Zhang et al. 2010); $continuous line$: tentative fit with a synthetic spectrum (see text); $dashed\, line$: contribution of ``continuum carriers, as opposed to ``band" carriers i.e. those carrying specifically the 21 and 30 $\mu$m bands (see Table 1). The standard deviation between spectrum and fit is 0.1 (in the same arbitrary emission units as the ordinates) i.e. less than 10 \% of the peak emission. Color on line.} 
\label{Fig:iras2}
\end{figure}

The general trends of the spectral contributions of the ``continuum" carriers in Fig. \ref{Fig:iras1} and \ref{Fig:iras2} are reminiscent of their counterparts in Zhang et al. \cite{zha10} except that our ``continua" are quite bumpy. Part of this bumpiness may really be associated with HAC structures, but another part should be cured by refining our ``HAC" models and concatenating many more spectra of similar model structures. These imperfect fits, as they stand, are only meant to stress that simple carbonaceous structures, such as those used here for fitting spectra, may help understand the composition of continuum-carrying dust.

\section{Conclusion}
The main outcomes of this work are as follows.

1) Several generic, aliphatic and aromatic, hydrocarbon structures were built by means of molecular modeling codes. Their vibrational spectra were obtained by Normal Mode Analysis. The molecules are large enough that their spectra extend up to several thousand micrometers. 

2) The skeletal or phonon modes are clearly identified by the distribution of the atomic velocity vectors in each mode. In all samples, their minimum wavelength is about 18 $\mu$m, as predicted by Debye's law. In spectra of pure hydrocarbons, the phonon region is clearly separated from the MIR fingerprints.

3) In every elementary frequency interval over the spectra, the dispersion of integrated absorption line intensities is considerable: about a factor 10. This is characteristic of disordered and ``impure" structures by contrast with pure, infinite crystals. As the particle size increases and/or as the number of slightly different structures increases along the same line of sight, the ``weak" modes in the MIR fingerprint regions form plateaus beneath the bands, and a continuum in the phonon region. No continuum is found below 6 $\mu$m. 

The fact that the same structure carries the fingerprints as well as their underlying plateaus 
validates the suggestion of Kahanp$\ddot{\rm a}\ddot{\rm a}$ et al. \cite{kah} and Kwok and Zhang \cite{kwo11} referred to in the Introduction. The same structure also carries part of the underlying continuum.

4)\bf The structures considered here to illustrate and explain the formation of plateaus and continua are inspired from the wide variety of constituents of kerogens taken as a whole. In space they may perhaps be invoked as precursors or debris of the proposed kerogen-like UIB carriers (Papoular 2011). Nevertheless, the underlying interpretation of plateau and continua are largely independent of this particular choice.\rm

5) Intensities all over the spectrum are considerably enhanced by the inclusion of only a few percent heteroatoms, especially oxygen. The absorptivities of our structures then become comparable to those of candidate model dust materials measured in the laboratory.

6) Computed absorption cross-sections per H atom in space are compatible with those deduced from extinction measurements in the solar neighborhood.

7) Increasing the size of a molecule at constant structure and composition does not change the general characters of its spectrum; only the spectral density of lines increases and the spectrum extends farther into the FIR. There is a smooth transition between spectra of molecules and grains.

8) This information was applied to the synthesis of spectra that fit observed PPN spectra. 
This exercise confirmed that simple carbonaceous structures may well provide at least part of the continuum underlying FIR features. It also suggested that the temperature of the continuum carriers are continuously distributed rather than limited to one or two specific values. 

These results may help understand the emergence of plateaus, the origin of continua underlying FIR bands, as well as the composition of circumstellar dust. 

9) The same interpretation should be applicable to plateaus and continua underlying the UIBs (3 to 20 $\mu$m). For instance, the slowly rising continuum of the cool dust component, at 80 K, in Fig. \ref{Fig:iras1} is reminiscent of the slowly rising spectrum of NGC 1482 (Smith et al. 2007) or the Orion Bar (Kwok and Zhang 2011). Indeed, UIBs are generally observed in colder, interstellar media, and so, both the composition and temperature of the accompanying ``dust" should be different than is the case for PPNe. The treatment of this case 
is outside the scope of the present paper.

\section{Acknowledgments}
I am grateful to Prof. Sun Kwok for providing the data files of the PPN spectra. I thank the anonymous referee for several comments and suggestions which helped improve the manuscript.

\end{document}